# Tracking interfacial changes of graphene/Ge(110) during in-vacuum annealing

L. Camilli[1,*], M. Galbiati[2], L. Di Gaspare[3], M. De Seta[3,*], I. Píš[4], F. Bondino[4], A. Caporale[3], V.-P. Veigang-Radulescu[5], V. Babenko[5], S. Hofmann[5], , A. Sodo[3], R. Gunnella[6] and L. Persichetti[1,3,*]

[1]Dipartimento di Fisica, Università di Roma "Tor Vergata", Via Della Ricerca Scientifica, 1- 00133 Rome, Italy

[2]Department of Physics, Technical University of Denmark, 2800 Lyngby, Denmark

[3]Dipartimento di Scienze, Università Roma Tre, Viale G. Marconi, 446- 00146 Rome, Italy

[4]IOM-CNR Laboratorio TASC Trieste I-34149, Italy

[5]Department of Engineering, University of Cambridge, Cambridge CB3 0FA, United Kingdom

[6]School of Science and Technology Physics Division, University of Camerino, Italy

Graphene quality indicators obtained by Raman spectroscopy have been correlated to the structural changes of the graphene/germanium interface as a function of in-vacuum thermal annealing. Specifically, it was found that graphene becomes markedly defective at 650 °C. By combining scanning tunneling microscopy, X-ray photoelectron spectroscopy and Near Edge X-ray absorption fine structure spectroscopy, we concluded that these defects are due to the release of $H_2$ gas trapped at the graphene/germanium interface. The $H_2$ gas was produced following the transition from the as-grown hydrogen-termination of the Ge(110) surface to the emergence of surface reconstructions in the substrate. Interestingly, a complete self-healing process was observed in graphene upon annealing at 800 °C. The identified subtle interplay between the microscopic changes occurring at the graphene/germanium interface and graphene's defect density is integral to advancing the understanding of graphene growth directly on semiconductor substrates, controlled 2D-3D heterogeneous materials interfacing and integrated fabrication technology.

[*]corresponding thors:

Luca Camilli

Phone: +39 0672594123; E mail: Luca.Camilli@uniroma2.it (Luca Camilli)

Monica De Seta

Phone: +39 0657333430; E mail: monica.deseta@uniroma3.it (Monica De Seta)

Luca Persichetti

Phone: +39 0672594547; E mail: Luca.Persichetti@uniroma2.it (Luca Persichetti)

# 1. Introduction

Graphene gives rise to many unique properties and device concepts [1-3], with transformational opportunities particularly for integrated solid state device technology [4, 5]. A current challenge across many technological roadmaps is 2D/3D materials and process interfacing, particularly achieving compatibility with scalable complementary metal-oxide-semiconductor (CMOS) technology [6-8]. Prevalent chemical vapor deposition (CVD) routes for graphene films rely on catalytic enhancement via transitional metal substrates, such as Cu [9], which requires the subsequent transfer and raises issues of metal contamination [10]. In this context, germanium (Ge) has emerged as a promising alternative growth substrate for direct CVD growth of graphene [11-14]. The lack of stability of Ge carbides together with the low carbon solubility in Ge enables the controlled synthesis of graphene on the Ge(001), Ge(110) and Ge(111) surfaces [15-20]. However, on Ge(001), the growth of high-quality graphene is typically accompanied by the faceting of the Ge substrate along shallow crystal orientations vicinal to the (001) face [14, 21], likely triggered by the well-known existence of several surface-energy minima around Ge(001) [22]. This is not the case for Ge(110), on which planarity is preserved during graphene growth and where wafer-scale single-crystal graphene can be achieved [11, 23].

Recent investigations have revealed a rich and complex scenario for the interface structure of graphene/Ge(110) [24-28]. As-grown CVD graphene samples feature a hydrogen-passivated Ge surface (α phase) which is stabilized under the $H_2$-rich growth conditions [11]. In-vacuum annealing triggers the desorption of hydrogen and the appearance of novel surface reconstructions which are not observed on bare Ge(110). Above 350 °C, the Ge(110) surface starts to reconstruct forming a large (6×2) surface cell (β phase) [29], while, above 700 °C, a different reconstruction is formed which shares structural motifs of the (1×1) Ge termination (γ phase) [26, 28]. The three phases may coexist on the Ge(110) surface under graphene at intermediate temperatures. The phase transitions imply relevant changes in surface termination and an extensive rearrangement of Ge surface atoms; this is expected to affect the graphene layer above. For instance, it has been shown that the electronic properties of graphene are influenced by these structural changes occurring on the Ge(110) surface [28]. To date, however, little is known about if and how these structural modifications of the graphene/Ge(110) interface influence the quality of the graphene adlayer. This is particularly pertinent to the many post-growth processing and fabrication steps of graphene involving thermal annealing.

In this paper, we systematically correlate the quality features of graphene/Ge(110) obtained by Raman spectroscopy as a function of post-growth in-vacuum annealing to the structural changes of the interface investigated by combining scanning tunneling microscopy (STM), synchrotron-radiation-excited X-Ray Photoelectron Spectroscopy (XPS) and Near Edge X-ray Absorption Fine Structure Spectroscopy (NEXAFS). Following thermal desorption of hydrogen from the Ge(110) surface and the emergence of (6×2)-reconstructed β patches on germanium, at 450 °C we observe characteristic spectral changes in the C 1s XPS peak of graphene which we correlate to the formation of graphene nanobubbles observed by STM and attributed to trapped hydrogen gas underneath graphene. Further annealing at 650 °C produces a dramatic increase of structural defects in graphene indicated by Raman spectroscopy which we ascribe to the release of the trapped gas. We find that these defects can be completely healed out by annealing to $T > 750$ °C which, in turn, results in the formation of a monodomain of the γ (1×1)-reconstructed phase on the Ge(110) surface. Promotion of defect healing during in-vacuum high temperature annealing [30] indicates the key role played by the catalytic activity of the Ge substrate, similarly to what has been observed for graphene on catalytic substrates like Cu [31, 32].

Despite the changes in graphene XPS and Raman spectra, NEXAFS reveals no hint of orbital mixing between graphene and Ge-derived states or any $sp^2$-to-$sp^3$ hybridization. Our findings reveal a subtle interplay between graphene quality and interface structural changes produced by post-growth thermal treatments in graphene/Ge(110) and could be especially useful for metal-free graphene integration pathways for emerging optoelectronic applications.

## 2. Materials and methods

Graphene films were grown on undoped Ge(110) substrates using a tube furnace CVD reactor with a base pressure of ~$2 \times 10^{-7}$ mbar. A gas mixture of $CH_4/H_2/Ar$ (1/100/1600 sccm) was employed at 800 mbar total pressure at ~925 °C. The temperature ramp-up occurs following three stages: a ramp-up to 875 °C in 40 minutes, followed by a slower rate ramp-up to 905 °C in 20 minutes and finally to 925 °C in 20 minutes. The cooling down from the growth temperature to 300 °C takes place in about 20 minutes. At this point, cooling to room temperature requires additional 2.5 hours. Samples were cooled down to room temperature in $H_2$ and Ar atmosphere. The samples were stored under vacuum and outgassed at around 100 °C for several hours in ultra-high-vacuum (UHV) conditions before STM and the photoelectron spectroscopy experiments. Thermal annealing treatments were performed in UHV. The annealing temperature was measured using a thermocouple/ optical pyrometer and the annealing duration was 40 minutes from the moment the peak temperature was reached.

STM measurements were performed with etched W tips at a temperature of 9-10 K in UHV (base pressure <$2\times10^{-10}$ mbar) using a Scienta Omicron Infinity System. Synchrotron radiation XPS, and NEXAFS measurements were performed at the BACH beamline [33, 34] of the Elettra synchrotron facility in Trieste, Italy. A VG Scienta R3000 hemispherical electron analyzer placed at 60° from the incident beam direction was used. XPS and NEXAFS measurements were performed at room temperature using light linearly polarized in the horizontal plane and spectra at different take-off angles were acquired by rotating the sample. C 1s and Ge 3d core-level spectra were recorded at a photon energy of 400 eV with a total spectral resolution of 170 meV. Note, however, that one should separate the spectral resolution from the capability of the system to detect energy shifts of individual peaks. Relative peak shifts can be detected with a much higher accuracy using peak fitting algorithms. Moreover, we recorded reference Au $4f_{7/2}$ core level peak measured on a gold sample regularly to compensate any possible monochromator (excitation photon energy) drifts. Thus, we could reliably detect core-level shifts as small as a few tens of meV. Peak fitting was performed using the KolXPD software after performing Shirley background subtraction. The NEXAFS spectra were acquired in partial electron yield mode recording C KVV Auger electrons using the hemispherical analyzer operated at 100 eV pass energy and fixed at a kinetic energy of 263 eV. Photoemission features were removed from the carbon Auger region using peak fitting following the procedure introduced by O. Lytken [35]. The NEXAFS intensities were normalized to the photon flux derived from the total photoelectric current from a clean gold mesh placed in the incoming beam. The photon energy resolution was set to 80 meV. Raman measurements were carried out at room temperature by using a Renishaw Raman InVia Reflex µ–spectrometer equipped with a diode laser at 532 nm, operating at a nominal output power of 100 mW, and with a Leica DM2700 M confocal microscope. A 1800 line/mm grating disperses the backscattered light and a Peltier cooled (-70°C) 1024×256-pixel CCD detector collects the Raman signal. The 520.5 cm$^{-1}$ line of the inner silicon standard sample is used to calibrate the spectra. WiRE™ 5.3 software licensed by Renishaw is used for measurements set-up, acquisition, subtraction of systematic errors, baseline and data smoothing. The spectra were recorded by using the 100× objective, with a nominal spectral resolution of about 1 cm$^{-1}$. The laser power on the surface of the samples, the integration time and the number of accumulations, are optimised in order to enhance the signal-to-noise ratio.

In order to obtain combined information from XPS, Raman and STM, annealing treatments reaching different maximum temperature setpoints were performed on different samples grown in the same conditions, so that each phase evidenced by synchrotron-radiation photoemission measurements was also characterized by STM and Raman.

3. Results

Figure 1 shows XPS spectra, collected at an angle of 60° with respect to the surface normal as a function of the annealing temperature for C 1s (upper panels) and Ge 3d core levels (lower panels). The lowest temperature corresponds to a mild outgas of the samples to remove adventitious contaminations adsorbed on the sample surface. All C 1s spectra show a main component fitted with Gaussian-broadened Donjach-Sunjic lineshape with an asymmetry parameter of 0.1 (yellow curve, labelled as *C*) associated to graphene. Between 100 °C and 650 °C, this component is centered at $284.3^{+0.01}_{-0.03}$ eV, while it shifts to 284.5 eV at 800 °C. The upshift could be possibly related to the decreased distance between graphene and germanium in the γ phase after high-temperature annealing as well as to the n-doping of graphene observed in this phase [28]. At the lowest temperature, we observe an additional broad Gaussian component with small amplitude located at +0.5 eV from the main graphene peak (highlighted in blue and labelled as $C_1$) which is typically associated to C-H and C-C sp$^3$ bonds originating from the absorption or trapping of CH$_4$ during the CVD growth process [36]. Another component, displayed in red and labelled as $C_2$, is centered at -0.39 eV with respect to the main graphene peak. It grows significantly in amplitude between 100 °C and 450 °C, while its amplitude again decreases at 650°C, finally vanishing at 800 °C. This $C_2$ component is well fitted by a Doniach-Sunjic lineshape with the same asymmetry and width as the main peak, suggesting that its origin can be related to a portion of C atoms in the graphene layer rather than to a different chemical bonding. For example, the appearance of low-binding energy components has been observed for C atoms in graphene "lifted up" by gas molecules underneath, such as CO, O, H [37, 38], or for C atoms in the graphene layer being subjected to modified core hole screening [39].

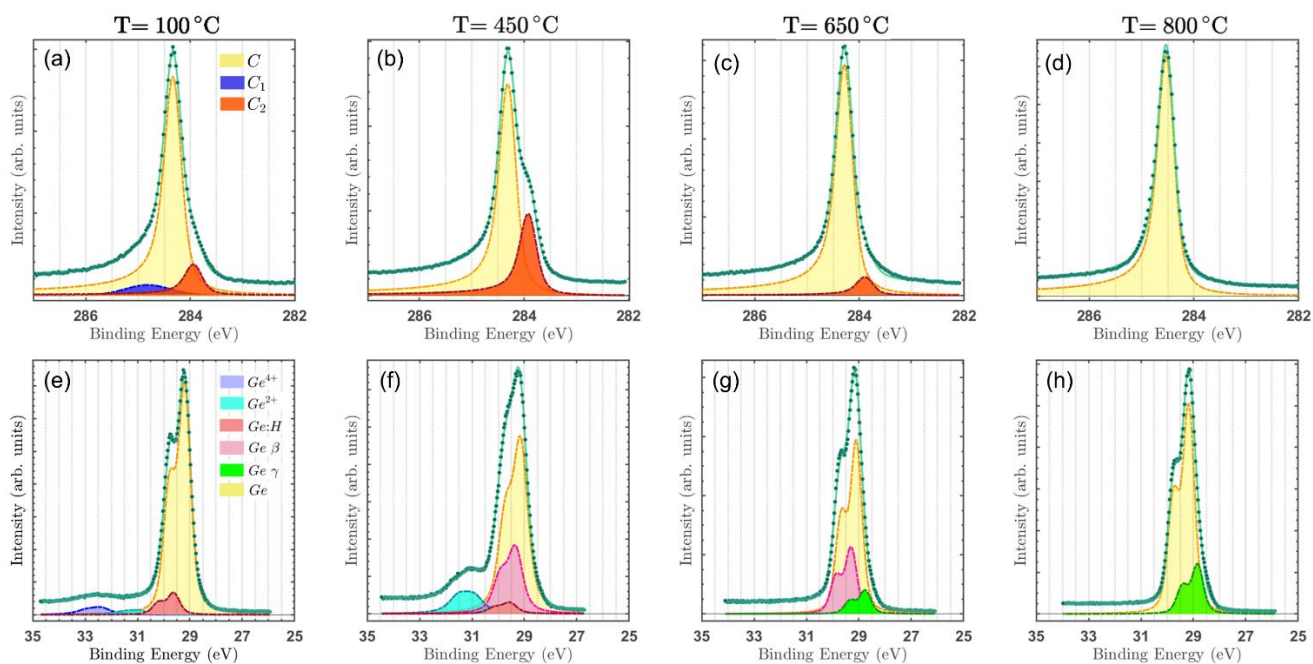

*Fig. 1. XPS spectra measured at room temperature with a photon energy of 400 eV at emission angle 60° as a function of the annealing temperature for C 1s (upper panels) and Ge 3d core levels (lower panels). Experimental datapoints are shown as green dots; green continuous lines are the fits to the experimental data obtained with the components displayed as colored shaded areas. The peak energy evolution of the fitting components with temperature for C 1s and Ge 3d is reported in Fig. S1 of Supplementary Information.*

The lower panels in Fig. 1 show the evolution with temperature of the Ge 3d XPS peak. The characteristic $3d_{5/2}$ and $3d_{3/2}$ doublet of elemental germanium (distance in energy being 0.58 eV) is fitted to a Voigt profile for which the ratio of the Gaussian and Lorentzian functions and the spin-orbit splitting were kept fixed (yellow curves labelled as *Ge*). The linewidth of this yellow *Ge* component is larger at 450 °C than at the other temperatures for which it remains roughly constant.

The red component labelled as *Ge:H* in Ge 3d core levels, visible at 100 °C and 450 °C and shifted by about 0.4 eV towards higher binding energy with respect to the main Ge peak, can be ascribed to the Ge surface atoms terminated by H with the formation of H-Ge bonds [25], in line with a H-passivation model proposed by Lee e*t al.* for as-grown CVD graphene/Ge(110) [11].

At 100 °C, we observe small components at binding energies corresponding to the $Ge^{4+}$ (dark blue) and the $Ge^{2+}$ (light blue) oxidation states, which we associate to the oxidized surface of germanium in the regions not covered by graphene. Besides, the germanium surface is likely not fully protected against oxidation at the boundary regions between graphene grains where oxygen could locally oxidize the germanium surface, penetrating from the graphene edges. Apart from these areas, the germanium surface fully covered by graphene should not be oxidized, as the graphene overlayer is known to provide a viable protection against the oxidation of Ge(110) at ambient pressure up to a year [40-43]. Concerning the germanium oxide, it is known that the native oxide layer consists of an interfacial $GeO_x$ layer (with dominant GeO stoichiometry) on which $GeO_2$ grows over time when the surface is exposed to air [41]. In our data, the amplitude ratio of $Ge^{2+}/ Ge^{4+}$ is about 0.6 at 100 °C, while at 450 °C the $Ge^{2+}$ is the only visible component. This can be explained, in part, by the attenuation of the photoelectrons originated in the buried GeO layer and absorbed by the overlying $GeO_2$ layer and, in part, by a progressive reduction of $Ge^{4+}$ to $Ge^{2+}$ during the annealing process in vacuum [44].

At 450 °C, we also observe a prominent component (evidenced in pink and labelled as *Ge β*) again at higher binding energy with respect to the bulk Ge XPS peak but shifted by about -0.2 eV with respect to the spectral feature of the H-passivated Ge surface. The relative energy position of the peak is compatible with the formation of the (6×2)-reconstruction on the Ge(110) surface [25]. For annealing to 650 °C, this component is still present and a new one shows up at -0.34 eV with respect to the bulk peak (bright green labelled as *Ge γ*), while the red component associated to the H-terminated Ge surface is lost. As it will be discussed in the next section, based on the analysis of STM data, we associate the bright green component to the γ surface reconstruction of germanium.

Finally at 800 °C, only the bright green *Ge γ* spectral feature is present together with the bulk *Ge* one.

The same components and their evolution with temperature are observed at near-normal emission geometry but, as expected, the relative amplitudes of the Ge surface features are diminished (See Fig. S2 in Supplementary Information).

We notice that the same evolution for C 1s spectra as in Fig. 1 has been also observed in a sample exposed to air for about one year where the initial oxidation of the germanium substrate is more severe (Fig. S3 in Supplementary Information), thus confirming that the oxide features in the Ge 3d peaks are not directly related to the changes occurring in graphene.

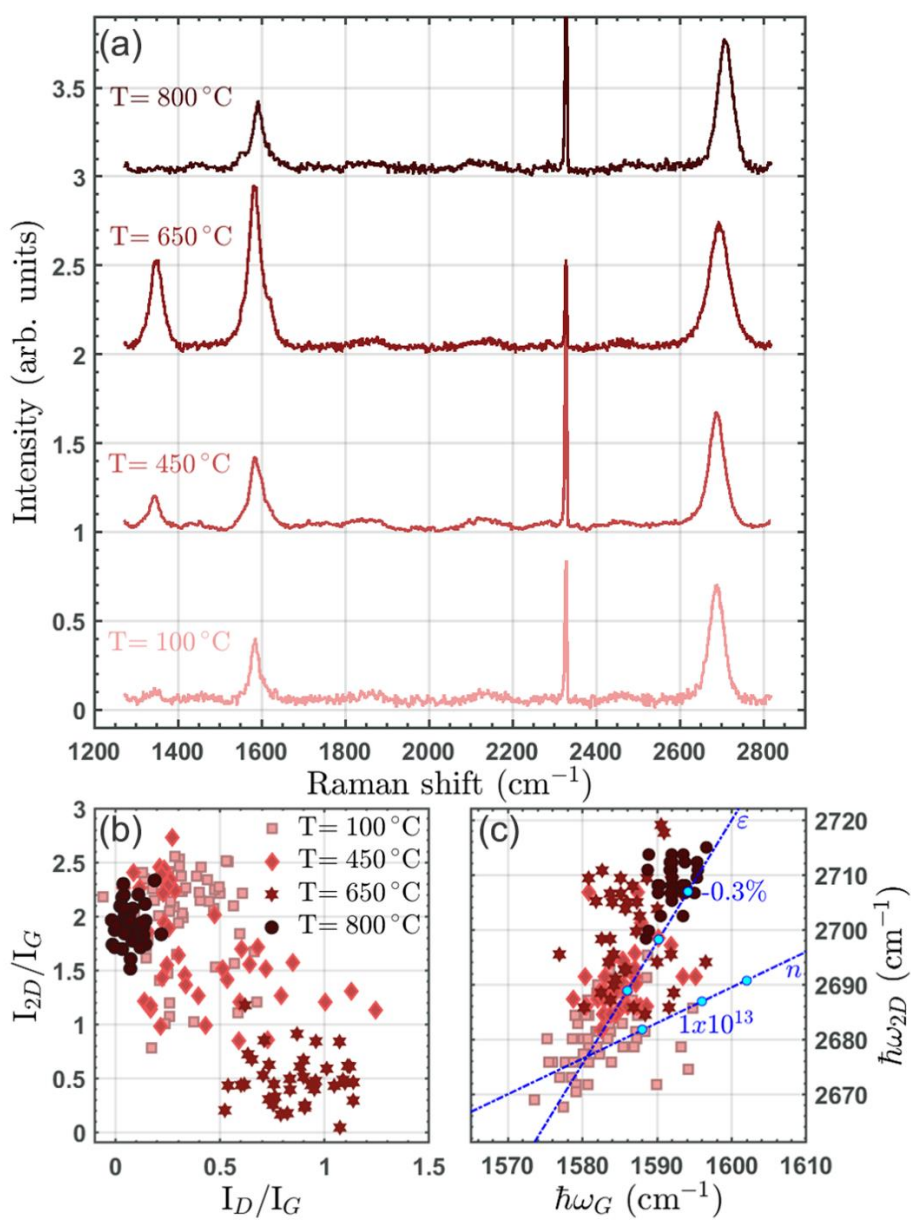

*Fig. 2. (a) Raman spectra of graphene/Ge(110) as a function of the annealing temperature acquired in air (the characteristic $N_2$ peak around 2330 cm$^{-1}$ is visible). (b) Correlation of the ratio of the 2D mode to G mode intensity (obtained from peak heights) versus the ratio of the D mode to the G mode intensity. (c) Plot of the 2D- vs G-mode energies. ε is the strain and n the charge density. Blue lines indicate $\hbar\omega_{2D}$ vs $\hbar\omega_G$ relationship for strained undoped (n= 0) and unstrained (ε= 0) doped graphene. The two lines cross at the expected 2D and G mode energies for undoped freestanding single-layer graphene.*

Figure 2 reports the Raman spectroscopy characterization of each of the annealing stages probed by XPS. Data reported in the figure were acquired on the same sample, on which we checked by XPS the presence at 450 °C of the peculiar shoulder shifted of -0.39 eV with respect to the main C 1s peak (see Fig S4 in Supplementary Information). We also checked for the overall consistency over different samples of the behavior observed by Raman as a function of temperature (Fig. S5 in Supplementary Information). Data are collected from mapping the samples over areas of about 200 µm$^2$ with a step of ~2 µm (Fig. S6 in Supplementary Information). The Raman spectra averaged out from the maps at the different annealing temperatures are displayed in Fig. 2(a). The spectral range includes the main peaks related to graphene, namely the 2D and G bands, which always satisfy the Raman selection rule [45], and, depending on temperature, are observed in our samples within the range 2686-2705 cm$^{-1}$ and 1580-1592 cm$^{-1}$, respectively. In addition, at 450 °C and, above all, at 650 °C, the Raman-forbidden D and D' bands (centered, respectively, at ~1345 cm$^{-1}$ and 1615 cm$^{-1}$) also appear in the spectrum. As well-known, the D (D') mode is activated by a single-phonon intervalley (intravalley) scattering process promoted by the presence of defects in graphene [46].

| T (°C) | $\hbar\omega_{2D}$ (cm$^{-1}$) | $\hbar\omega_G$ (cm$^{-1}$) | $\Gamma_{2D}$ (cm$^{-1}$) | $\tilde{I}_{2D}/\tilde{I}_G$ | $\tilde{I}_D/\tilde{I}_G$ |
|---|---|---|---|---|---|
| 100 | 2689 | 1585 | 40 | 3.0±0.5 | 0.20±0.10 |
| 450 | 2687 | 1583 | 44 | 2.1±0.8 | 0.30±0.17 |
| 650 | 2693 | 1584 | 59 | 1.3±0.4 | 0.60±0.20 |
| 800 | 2705 | 1592 | 42 | 2.8±0.4 | <0.10±0.05 |

*Table 1. Quantitative analysis of Raman spectra. Data are obtained from the average spectra of Raman maps at each temperature reported in Fig. 2(a). The Raman ratios in the table are obtained from the integrated intensity of individual modes fitted to a Lorentzian lineshape. The uncertainties are estimated as the variance of the individual data in the maps.*

In Table 1, we report relevant parameters obtained from the average spectra by fitting the individual modes to a Lorentzian line shape, including the ratios of the integrated intensity of the 2D peak to that of the G peak ($\tilde{I}_{2D}/\tilde{I}_G$) and of the D peak over the G peak ($\tilde{I}_D/\tilde{I}_G$). To decouple the information related to the amplitude and width of the peaks, we also display in Fig. 2(b) the distribution over the maps of the corresponding ratios obtained from peak heights. By analyzing the evolution of the spectral fingerprints with temperature [Fig. 2(a)], we observe that at 100 °C, the D peak is barely visible, this highlighting the high quality of the as-grown samples. Also, the 2D band is sharp and can be fit with a single Lorentzian peak with a full-width-at-half-maximum (FWHM), $\Gamma_{2D}$, of 40 cm$^{-1}$. The integrated intensity of the 2D peak over the G peak ($\tilde{I}_{2D}/\tilde{I}_G$) is ~3.0, as expected for monolayer graphene on germanium [18, 23], while that of the D peak over the G peak ($\tilde{I}_D/\tilde{I}_G$) is ~0.2. At 450 °C, the $\tilde{I}_{2D}/\tilde{I}_G$ ratio slightly decreases while both $\Gamma_{2D}$ and $\tilde{I}_D/\tilde{I}_G$ exhibit a small increase (Table 1). At this temperature, Figure 2(b) shows that data are most spread out, suggesting that some spatial inhomogeneity affects the graphene layer. It is however at 650 °C that we observe the most striking changes. Annealing to this temperature results in a drastic increase in the amplitude of both the D and G peaks with respect to the 2D one, the relative amplitude of the D peak compared to the G one being also higher [Figs. 2(a) and 2(b)]. The D' mode is now clearly visible. The ratios obtained from the integrated intensities confirm the picture, while also highlighting the changes occurring in the peak widths. In particular, $\Gamma_{2D}$ increases by about 47% between 100 °C and 650 °C. This together with the drop of $\tilde{I}_{2D}/\tilde{I}_G$ indicate a lower structural quality of graphene at 650 °C, where the higher $\tilde{I}_D/\tilde{I}_G$ ratio shows that a large number of defects is introduced [47]. At 650 °C, the $\tilde{I}_D/\tilde{I}_{D'}$ is about 7 (see Fig. S7 in Supplementary Information), suggesting that the majority of these defects are vacancies [48].

The defected graphene becomes effectively self-healed by UHV annealing to about 800 °C. As evident in Fig. 2(a), at this temperature the D and D' peaks completely vanish and the structural and crystalline quality of graphene is restored, as further indicated by the decrease of the amplitude of the G mode with respect to the 2D one [Fig. 2(b)]. From Table 1, we see that the original $\tilde{I}_{2D}/\tilde{I}_G$ ratio is in fact recovered, as well as the FWHM of the 2D peak. It is interesting to note that, after annealing to 800 °C, even the tiny the Raman D peak present at 100 °C is completely absent, as in the case of spectra measured immediately after the sample growth [Fig. S8 in Supplementary Information].

From Table 1, it is also evident a shift of Raman peaks with temperature. In order to understand the origin of such shift, we followed the method proposed by Lee *et al*. [49] analyzing the evolution of Raman peak positions with temperature and correlating them to variations of strain and charge doping using the $\hbar\omega_G$-$\hbar\omega_{2D}$ space and suspended monolayer graphene as the origin [Fig. 2(c)]. From 100 °C to 800 °C, datapoints remain approximately parallel to the strain axis, reflecting mostly variation in the graphene strain (while the charge doping is likely always dominated by adsorbates on graphene, since Raman is measured in air). In line with the seminal work by Kiraly *et al*. [12], the native compressive strain, which is typically observed in CVD-grown graphene on Ge(110) [19], becomes larger as temperature is increased, evidencing that some changes are occurring in the interfacial properties of graphene on the Ge(110) surface as a function of temperature.

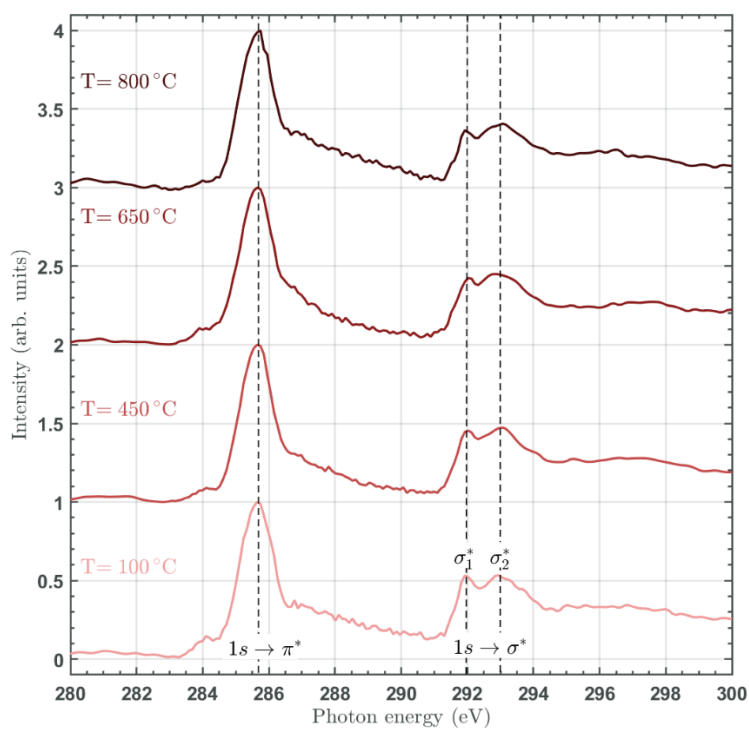

*__Fig. 3__. NEXAFS spectra at the C K-edge measured with the beam at 54.7° incidence with respect to the sample surface as a function of the annealing treatments of the graphene/Ge(110) system.*

Figure 3 shows the results of NEXAFS measurements at the C *K*-edge as a function of annealing treatments of the graphene/Ge(110) system to the same temperatures discussed by XPS and Raman. By scanning the photon energy around the value of the binding energy of the C 1s electrons, NEXAFS probes the resonances due to the excitation of C 1s electrons to unoccupied valence-band states of graphene [50]. The spectra in Fig. 3 are taken with the beam at 54.7° incidence with respect to the sample surface. At this "magic angle", the resonances appear in the spectrum independent of the orientation of the corresponding orbital [51]. We thus observe both the features due to the transitions from C 1s core levels to the partially occupied or empty π- and to the σ- states, namely the π* resonance at 285.6 eV and the σ* resonances at 292 eV and 293 eV [50, 52]. As evident in Fig. S9, the π* resonance is enhanced in grazing incidence spectra, i.e., that is with the polarization vector of the incident X-rays perpendicular to the basal plane of graphene. While the σ* resonances are enhanced in normal incidence spectra when the X-ray polarization vector is parallel to graphene. The 1s → σ* features a double peak structure. The first peak $\sigma_1$* is assigned to excitonic states and the second peak $\sigma_2$* is related to band-like contributions [52]. The intensity, shape, and position of the π* resonance are sensitive to the π-bond order and chemical environment. For example, if a C atom is bonded to an atom with higher electronegativity, it will make the C atom more electropositive than a C bonded to another C. The different bonding would affect the orbital energy and, in this case, the π* transition in the spectrum would be shifted to higher photon energies. Conversely, a shift of the π* transition to lower photon energy would be expected in the case of C atoms bonded with atoms with lower electronegativity. However, figure 3 shows no such features. Neither the separation between the 1s → σ* and the 1s → π* lines is changed with temperature. These observations rule out at any temperature the formation of strong bonds between C and Ge and the occurrence of a partial sp$^3$ C hybridization. Besides, the region between the π* and σ* lines does not show any additional peaks, indicating a negligible amount of functional groups bonded to the graphene layer [53]. We notice a smoothing of the splitting of the σ* resonance upon annealing at 650 °C, which is compatible with the presence of defects in the graphene structure [50, 54] at this temperature, as evidenced by Raman measurements. Looking at the region below the π* resonance, one can notice a small pre-edge peak centered around 284 eV, which vanishes almost completely at 800 °C. The origin of this feature had been originally attributed to the unoccupied density of states of single-layer graphene [50]. However, more recent calculations showed that such state should not be present in ideal free-standing graphene [55]. Indeed, also for the much more investigated growth of graphene on metals, the origin of such spectral feature is largely debated, with very different explanations proposed for its appearance. For example, the different degree of interaction between graphene and its substrate [56, 57] has been shown to strongly affect the pre-peak, which is observed on graphene on Pt(111) (where graphene is very weakly chemisorbed), it is shallower on graphene on Ir(111) (still weak chemisorption but slightly stronger than on Pt) whereas it is not observed on graphene on Ni(111) (case of strong chemisorption) [57]. On the other end, other authors correlated the NEXAFS pre-peak to modifications of the graphene electronic structures produced by structural defects [55]. It has been also suggested that the same peak could be correlated with localized electron doping of graphene, in particular in the case of an inhomogeneous distribution of discrete doped and undoped graphene regions created by charge transfer [58]. In our specific case, all these effects could combine in a complex way depending on temperature. As a matter of fact, the changes observed in the interface features (desorption of intercalated hydrogen, appearance of different, diverse reconstruction terminations of the Ge surface) affect the graphene/Ge interaction and, at the same time, structural defect abundancy changes with temperature. Thus, the attribution of this spectral feature to a specific individual effect is difficult. Nonetheless, our experimental observations, showing that on the same sample, the NEXAFS pre-peak at 284 eV does or does not appear depending on the thermal annealing treatment driving structural modifications of the Ge substrate and of the Ge/graphene interface, appear to be consistent with the understanding that such NEXAFS spectral feature is not intrinsically related to the electronic structure of ideal free-standing graphene.

## 4. Discussion

The changes in the quality and interfacial properties of graphene/Ge(110) driven by post-growth annealing in UHV and observed by XPS, Raman and NEXAFS spectroscopy will be now complemented and discussed in light of the morphological and structural information gained by STM experiments.

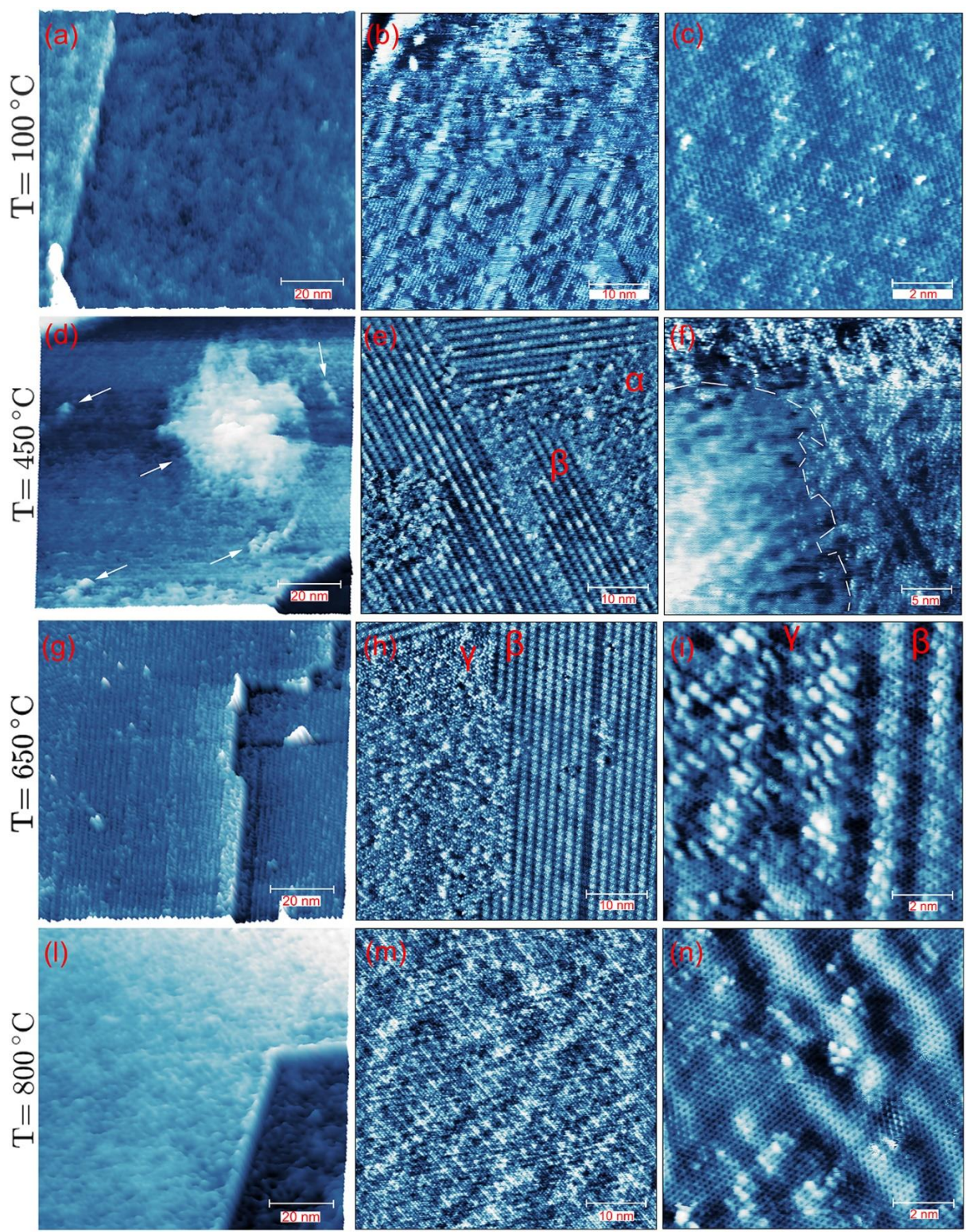

***Fig. 4****. (a) STM images measured at 10 K showing the graphene and Ge(110) surface as a function of the annealing temperatures. Each row corresponds to a different temperature. From left to right, the panels show increasing magnifications.*

Figure 4 shows STM images of graphene/Ge(110) obtained after annealing at increasing temperatures. At 100 °C (first row), while the mesoscale topography features - that is, monoatomic steps and flat terraces, of the Ge surface underneath graphene - are clearly visible [Fig. 4(a)], the atomic order of Ge(110) surface is barely accessible by STM [Fig. 4(b)] [12, 28], in line with the termination of the surface by hydrogen in the α phase [11] which was also hinted by our XPS data. The hydrogen passivation enables imaging the honeycomb lattice of graphene without strong cross-talks with the Ge surface states [Fig. 4(c)].

After annealing to 450 °C (second row), STM images show the formation of bubble-like protrusions on the Ge(110) terraces [indicated by arrows in Fig. 4(d)]. These structures have a variety of lateral sizes (see also Fig. S10 in Supplementary Information), some of them, like the largest one in Fig. 4(d), reaching several tens of nanometers. Their apparent height is below 1 Å, typically ranging between 200-800 pm. The formation of the bubbles, confirming our previous report in Ref. [28], is concurrent with the appearance of the (6×2) reconstruction (β phase) on the Ge(110) surface. Indeed, the enlarged view reported in Fig. 4(e) shows that β-phase patches coexist with other areas where the α phase is still locally present. The coexistence of these two phases on the Ge(110) surface at 450 °C well matches the presence of both the *Ge:H* (red) and *Ge β* (pink) spectral features in the Ge 3d XPS spectra in Fig. 1(f). Since the formation of the (6×2) reconstruction from the hydrogen-terminated germanium surface of the as-grown samples implies the rupture of H-Ge bonds [29], this transition produces hydrogen gas molecules that are likely to be trapped by the high-quality graphene monolayer [59]. A close-up of the bubble contour, highlighted in Fig. 4(f), shows that the features associated to the electronic states of the Ge substrate, which dominate the STM image outside the bubble, are not present inside it. Conversely, the graphene layer becomes visible in the bubbles' area, as evident in Figs. S10 and S11 in the Supplementary Information. This indicates that the physical separation between graphene and germanium is locally increased where the bubble is, and that the bubble itself is therefore enveloped by the graphene layer. As a matter of fact, the formation of graphene nanobubbles as a result of the local desorption of hydrogen atoms from the hydrogen terminated Ge(110) surface has been already observed in this system, either being triggered by an electrical [60] or thermal stimulus [28]. Our morphological characterization shows that, despite some bubbles might be already formed at lower temperature or still present at higher, their number and lateral size becomes striking at 450 °C. By looking back at the C 1s XPS spectra in Fig. 1, the most relevant feature distinguishing the *T*= 450 °C data from the others is the emergence of the red *$C_2$* component at lower binding energy with respect to the main graphene peak. We propose this component in C 1s to be related to trapping of hydrogen molecules underneath graphene. *Ab initio* studies indeed indicate that when hydrogen molecules are confined close to the graphene surface significant charge polarization effects occur, with negative charge accumulating on C atoms close to the $H_2$ molecule [12, 28, 61, 62]. The additional valence electron charge transferred to C atoms in the region of the bubbles is expected to modify the core hole screening and, in particular, to decrease the electrostatic potential at the core of the C atoms, thus lowering the energy necessary to remove their core electrons [39, 63, 64]. This initial state effect, compatible with the presence of $H_2$ inside the graphene nanobubbles, would therefore introduce an asymmetry of the C 1s peak to lower binding energy, which is indeed what we observe. We propose that the uneven structural and bonding landscape resulting from the simultaneous presence on the Ge substrate of hydrogen-terminated α phase, (6×2)-reconstructed β phase and $H_2$-filled graphene bubbles explains the broadening of the Ge 3d peak at 450 °C. Such inhomogeneity of the Ge substrate at 450 °C could also explain the larger spatial inhomogeneity observed in Raman data at this temperature where the individual datapoints obtained from map are the most spread out [Fig. 2(b)] and show larger variance (Table 1).

At 650 °C, the STM investigation reveals that the large graphene bubbles are no more present [Fig. 4(g)] and, on the Ge substrate, domains with a regular pattern of the γ reconstruction [26, 28] now coexist with the (6×2)-reconstructed β areas [Fig. 4(h)]. The progressive stabilization of the γ phase during high-temperature annealing has been previously demonstrated [28]. From the Ge 3d data in Fig. 1(g), we find that appearance of the γ phase corresponds to the show-up of the bright-green *Ge γ* surface component at low binding energy, which we therefore attributed to the γ reconstruction of the Ge(110) surface. On the other hand, the emergence of a fully reconstructed germanium surface, in a mixture of β and γ phases, is in line with the vanishing *Ge:H* red component of the α phase in Ge 3d at 650 °C [Fig. 1(g)]. The appearance of surface reconstructions on the germanium surface produces strong electronic crosstalk in STM between the graphene and the surface morphology of germanium, as evident in Fig. 4(i) where we image, at higher magnification, the boundary region between the β and γ areas. In the STM image, we observe, superimposed to the graphene honeycomb in the background, prominent features appearing as white protrusions following the symmetry of the Ge surface reconstructions underneath.

The disappearance of graphene bubbles at 650 °C means that the $H_2$ gas trapped below graphene had been released. We propose that its release produces the significant generation of defects in graphene observed by Raman at 650 °C (Fig. 2). The value of the $\tilde{I}_D/\tilde{I}_{D'}$ ratio in Raman at this temperature suggests that a significant amount of such defects are vacancies. Increased number of structural defects after post-growth hydrogen intercalation was observed on graphene/Ge(001) [65], with concurrent increase of the compressive strain. In our case, the observed increase of compressive strain with temperature does not appear related in a trivial way with defective graphene. As proposed by Kiraly *et al.* [12], it seems reasonable to link this increase to the interfacial changes resulting from the emergence of reconstructions on Ge(110) surface. Indeed, a further annealing of the samples to 800 °C produces a larger compressive strain [Fig. 2(c)] while the structural quality of graphene is simultaneously completely restored, as demonstrated by the Raman analysis in Fig. 2 and by the presence of a single sharp fitting component in the C 1s XPS peak at 800 °C [Fig. 1(d)]. At this temperature, STM data shows that γ is the only phase present on the Ge surface [Fig. 4(m)], consistently with the disappearance of the pink XPS component related to the β phase in Ge 3d [Fig. 1(h)]. STM imaging at higher magnification in Fig. 4(n) again shows the graphene honeycomb to which electronic features of the underneath germanium surface are overlapped. The simultaneous evolution of the Ge surface reconstruction towards the homogeneous γ phase and the recovery of the graphene quality suggests that the interaction with Ge surface may play a role in the graphene healing during the high temperature annealing. The influence of the underneath template for the structural reconstruction of graphene has been reported in Ref. [66]. On the other side, as observed by Chen et al. in exfoliated graphene on $SiO_2$ [30], at 800 °C displaced carbon atoms produced with the vacancy formation would acquire enough thermal energy to overcome the migration barrier and diffuse, promoting recombination with vacancies and their annihilation, and thus the healing of defects that we observe by Raman at 800 °C. Confirming previous results [30, 67], self-healing process promoted by in vacuum annealing is found to be very efficient in curing defected graphene without an external carbon source.

## 5. Conclusions

We investigated the impact of post-growth thermal annealing in vacuum on the quality of CVD-grown graphene on Ge(110). Our data reveal a non-trivial evolutionary pathway of Raman spectra with temperature, where graphene is markedly defective at around 650 °C, while an effective self-healing process occurs after annealing to 800 °C. This evolution is discussed and interpreted by employing XPS and NEXAFS spectroscopy, combined with microscopic information collected by STM. We trace back the formation of defects to the release of trapped hydrogen gas following the transition from the as-grown hydrogen-terminated surface of Ge(110) to the appearance of surface reconstructions in the substrate. Our results are relevant for developing widely required integrated graphene process pathways on conventional semiconductors and advancing the understanding of controlled 2D-3D heterogeneous materials interfacing.

**Acknowledgements**

The research leading to this result has been supported by the project CALIPSOplus under Grant Agreement 730872 from the EU Framework Programme for Research and Innovation HORIZON 2020. We acknowledge Elettra Sincrotrone Trieste for providing access to its synchrotron radiation facilities and the support from Lazio Innova (Regione Lazio, Italy) through the project "Gruppi di ricerca 2020" A0375-2020- 36566. V.-P. V.-R. and S.H. acknowledge funding from EPSRC (EP/M508007/1, EP/P005152/1) and support from NPL. L.C. acknowledges support from the 'Programma per Giovani Ricercatori'- Rita Levi Montalcini 2017. L.C. and M.G. acknowledge funding from the Villum Young Investigator Program (Project No. 19130). I.P. and F.B. also acknowledge funding from EUROFEL project (RoadMap Esfri).